\documentclass[12pt]{article}
\usepackage[cp1251]{inputenc}
\usepackage[english,russian]{babel}
\usepackage{cite}

\begin{document}

\title{Tomographic Representation of Quantum Mechanics and
Statistical Physics }

\author{Olga V. Man'ko\thanks{%
P. N. Lebedev Physical Institute, Leninskii Prospect 53, Moscow
119991, Russia, e-mail: omanko@sci.lebedev.ru}}

\date{}

\maketitle

\begin{abstract}
A review of the photon-number tomography and symplectic tomography
as examples of star-product quantization is presented. The classical
statistical mechanics is considered within the framework of the
tomographic representation.
\end{abstract}

\section{INTRODUCTION}

In quantum mechanics, the state is described by the wave function or
density matrix (density operator). In classical mechanics, the state
of system with fluctuations is described by the
probability-distribution function. The quantum and classical natures
are described using different objects -- operators and functions.
But for more easy understanding the nature of systems which has
classical and quantum parts, it is necessary to have the same
language for both fields. So, we need the representation in which
classical and quantum states are described by the same object -- the
probability distribution function. For this purpose, the tomographic
probability representation of quantum mechanics for continious
variables called symplectic tomography was introduced in
\cite{Mancini1,Mancini2,Mancini3}. The probability representation of
quantum mechanics was extended to the case of discrete variables in
infinite domain (photon-number tomography) in
\cite{vogel,wodk,Mancini3}. The name photon-number tomography was
given in \cite{Mancini3} for elucidating the physical meaning of
measuring the quantum state by means of measuring the number of
photons (i.e., photon statistics). Photon-number tomography is the
method to reconstruct the density operator of quantum state using
measurable probability-distribution function (photon statistics)
called tomogram. The tomographic-probability representation of
classical states was introduced in
\cite{OMVI,MankoManko1,Pilyavets,MankoManko}. In the probability
representation, the quantum states and the classical states are
described by the same objects -- tomograms. Tomograms are positive
measurable probability-distribution functions of random variables.
In order to describe observables by functions instead of operators
in quantum mechanics, the quantization based on star-product of
functions is used. In \cite{Marmo,Kapusta} it was shown that the
symplectic-tomography scheme is a new example of quantization based
on star-product of functions -- symbols of operators. In
\cite{KuzOVSPIE,127} it was shown that photon-number tomography is
an example of the star-product quantization.

The aim of this paper is to present a review of the tomographic
approaches and their connection with star-product quantization
scheme.

\section{GENERAL STAR-PRODUCT SCHEME}

Following \cite{Marmo,Kapusta}, we consider a density operator
$\hat \rho $ acting in a given Hilbert space. Let us suppose that we
have a set of operators $\hat{\cal U}({\bf x})$ acting in the
Hilbert space $H$, where the n-dimensional vector ${\bf
x}=(x_1,x_2\ldots,x_n)$ labels the particular operator in the set. We
construct the $c$-number function $w_{\hat\rho}({\bf x})$ using the
definition
\begin{equation}\label{Seq.1}
w_{\hat \rho}({\bf x})=\mbox{Tr}\,(\hat \rho\hat{\cal U}({\bf x})).
\end{equation}
The function $w_{\hat\rho}({\bf x})$ is called tomogram, it is the
symbol of density operator $\hat \rho$, and the operators $\hat{\cal
U}({\bf x})$ are dequantizers \cite{Patricia}. We suppose that
this relation has the inverse, i.e., there exists the set of operators
$\hat{\cal D}({\bf x})$ acting in the Hilbert space such that
\begin{equation}\label{Seq.2}
\hat \rho= \int w_{\hat \rho}({\bf x})\hat{\cal D}({\bf x})~d{\bf
x}.
\end{equation}
The operators $\hat{\cal D}({\bf x})$ are quantizers
\cite{Patricia}. Formulas (\ref{Seq.1}) and (\ref{Seq.2}) are
selfconsistent, if one has the following property of the quantizers
and dequantizers:
\begin{equation}\label{SVs15}
\mbox{Tr}\left[\hat{\cal U}({\bf x}) \hat{\cal D}({\bf x}')\right]=
\delta\left({\bf x}-{\bf x}'\right).
\end{equation}
Relations~(\ref{Seq.1}) and~(\ref{Seq.2}) determine the
invertable map of the density operator $\hat \rho$ onto function --
tomogram $w_{\hat \rho}({\bf x})$. We introduce the associative
product (star-product) of two tomograms $w_{\hat\rho_1 }({\bf x})$
and $w_{\hat\rho_2}({\bf x})$ corresponding to two density operators
$\hat\rho_1$ and $\hat\rho_2$, respectively, by the relations
\begin{equation}\label{eq.5}
w_{\hat\rho_1\hat\rho_2}({\bf x})=w_{\hat\rho_1}({\bf x})\ast
w_{\hat\rho_2} ({\bf x})=\mbox{Tr}\,(\hat\rho_1\hat\rho_2\hat{\cal
U}({\bf x})).
\end{equation}
The map provides the nonlocal product of two tomograms
(star-product)
\begin{eqnarray*}
w_{\hat\rho_1}({\bf x})\ast w_{\hat\rho_2}({\bf x})=\int w_{\hat
\rho_1}({\bf x}'')w_{\hat\rho_2}({\bf x}') K({\bf x}'',{\bf x}',{\bf
x})\,d{\bf x}'\,d{\bf x}''.
\end{eqnarray*}
The kernel of the star-product is linear with respect to the
dequantizer and nonlinear in the quantizer operator
\begin{eqnarray*}
K({\bf x}'',{\bf x}',{\bf x})= \mbox{Tr}\left[\hat{\cal D}({\bf
x}'')\hat{\cal D}({\bf x}') \hat{\cal U}({\bf x})\right].
\end{eqnarray*}
The associativity condition for tomograms means that the kernel of
star-product of tomograms $K({\bf x}'',{\bf x}',{\bf x})$ satisfies
the nonlinear equation \cite{Patricia}
\begin{equation}\label{patC2}
\int K({\bf x}_1,{\bf x}_2, {\bf y})K({\bf y},{\bf x}_3, {\bf
x}_4)d{\bf y}=\int K({\bf x}_1,{\bf y}, {\bf x}_4)K({\bf x}_2,{\bf
x}_3, {\bf y})d{\bf y}.
\end{equation}

\section{SYMPLECTIC TOMOGRAPHY}

Now we consider the symplectic-tomography scheme \cite{Mancini1} as an
example of the star-product quantization, following
\cite{Marmo,Kapusta}. In this scheme, the
tomographic symbol of the density operator $\hat\rho$ called
symplectic tomogram $w_{\hat \rho}({\bf x})$ is obtained due to
fomula (\ref{Seq.1}) by means of the dequantizer
\begin{eqnarray*}
\hat{\cal U}(X,\mu,\nu)=\delta(X\hat 1-\mu\hat q-\nu\hat p),
\end{eqnarray*}
where vector ${\bf x}=(X,\mu,\nu)$ has the coordinates which are
real numbers and $\hat 1$ is identity operator. The operators $\hat
q$ and $\hat p$ are the position and momentum operators,
respectively. The symplectic tomogram $w\,(X,\,\mu,\,\nu )$ depends
on the two extra real parameters $\mu $ and $\nu $. The physical
meaning of the parameters $\mu $ and $\nu $ is that they describe an
ensemble of rotated and scaled reference frames in which the
position $X$ is measured. For $\mu =\cos \,\varphi $ and $\nu =\sin
\,\varphi ,$ the symplectic tomogram coincides with the marginal
distribution for the homodyne-output variable used in optical
tomography~\cite{Ber,VogRi}.

Symplectic tomogram can be determined as the expectation value of the
delta-function calculated with the help of the density operator
\begin{equation}\label{eq.1b} w\left (X,\,\mu ,\,\nu \right )=
\langle \delta\left (\mu \hat q +\nu \hat p -X\right )\rangle.
\end{equation}
The symplectic tomogram is nonnegative function
\begin{eqnarray*}
w\left (X,\,\mu ,\,\nu \right )\geq 0,
\end{eqnarray*}
and it is normalized with respect to the variable $X$
\begin{eqnarray*}
\int w\left (X,\,\mu ,\,\nu \right )\,d X=1.
\end{eqnarray*}
The tomogram is a homogenious function
\begin{eqnarray*}
w\left (\lambda X,\,\lambda\mu ,\,\lambda\nu \right
)=|\lambda|^{-1}w\left (X,\,\mu ,\,\nu \right ).
\end{eqnarray*}
The density operator of the state can be reconstructed in the
symplectic tomography scheme, in view of (\ref{Seq.2}), with the
help of quantizer
\begin{eqnarray*}
\hat{\cal D}(X,\mu,\nu)=\frac{1}{2\pi} \exp\left(iX\hat1-i\nu\hat
p-i\mu\hat q\right).
\end{eqnarray*}
Since the density operator determines completely the quantum state
of a system and, on the other hand, the density operator itself is
completely determined by symplectic tomogram, one can use for
describing quantum states symplectic tomograms (positive and
normalized) which are probability-distribution functions analogous
to the classical ones. The quantum state is given if the position
probability distribution $w\left(X,\,\mu,\,\nu \right)$ in an
ensemble of rotated and squeezed reference frames in phase space is
given. The information contained in symplectic tomogram
$w\left(X,\,\mu,\,\nu \right)$ is overcomplete. To determine the
quantum state, it is enough to know the values of symplectic tomogram
of the state for arguments satisfying the following condition:
$\left(\mu^2+\nu^2=1\right),$ where $\mu=\cos \varphi$.

The kernel of star-product of two symplectic tomograms of density
operators $\hat\rho_1$ and $\hat\rho_2$  has the following form
\cite{Marmo}:
\begin{eqnarray*}
&&K(X_1,\mu_1,\nu_1,X_2,\mu_2,\nu_2,X\mu,\nu)=
\frac{\delta\Big(\mu(\nu_1+\nu_2)-\nu(\mu_1+\mu_2)\Big)}{4\pi^2}\nonumber\\
&&\times\exp\Big(\frac{i}{2}\Big\{\left(\nu_1\mu_2-\nu_2\mu_1\right)\Big.\Big.
\Big.\Big.+2X_1+2X_2-\frac{2(\nu_1+\nu_2)X}{\nu}\Big\}\Big).
\end{eqnarray*}

\section{PHOTON-NUMBER TOMOGRAPHY}

The photon-number tomogram defined by the relation
\begin{equation}\label{eq.1}
\omega(n,\alpha)=\langle n\mid\hat D(\alpha)\hat \rho\hat
D^{-1}(\alpha)\mid n\rangle
\end{equation}
is the function of integer photon number $n$ and complex number
$\alpha=\mbox{Re}\,\alpha+i\,\mbox{Im}\,\alpha,$ where $\hat\rho$ is
the state density operator and $\hat D(\alpha)$ is the Weyl
displacement operator \[\hat D(\alpha)=\exp(\alpha\hat
a^{\dagger}-\alpha^*\hat a).\] The photon-number tomogram is a
symbol of the density operator
\begin{equation}\label{eq.12c}
\omega(n,\alpha)=\mbox{Tr}\left[\hat\rho\,\hat{\cal U}(\mbox{\bf
x})\right],
\end{equation}
where dequantizer $\hat{\cal U}(\mbox{\bf x})$ reads \[\hat {\cal
U}(\mbox{\bf x})=\hat D(\alpha)|n\rangle\langle n|\hat
D^{-1}(\alpha), \,\mbox{\bf x}=(n,\alpha).\] The density operator
can be reconstructed from the photon-number tomogram with the help
of the inverse formula \cite{vogel,wodk,Mancini3}
\begin{equation}\label{eq.12b}
\hat\rho=\sum_{n=0}^{\infty}\int \frac{4\,d^2\alpha}{\pi(1-s^2)}
\left(\frac{s-1}{s+1}\right)^{(\hat a^\dagger+\alpha^*) (\hat
a+\alpha)-n}\omega(n,\alpha),
\end{equation}
where $s$ is an arbitrary ordering parameter~\cite{cahillglauber}.
The quantizer in the photon-number-tomography scheme is
\begin{equation}\label{eq.14}
\hat{\cal D}(\mbox{\bf x})=\frac{4}{\pi(1-s^2)}
\left(\frac{s-1}{s+1}\right)^{(\hat a^\dagger+\alpha^*)(\hat
a+\alpha)-n}.
\end{equation}
It is known \cite{cahillglauber} that the Wigner function
\cite{Wigner32}, which corresponds to the density operator
$\hat\rho$, is given by the expression
\begin{equation}\label{eq.2}
W_{\hat\rho}(q,p)= 2\,\mbox{Tr}\left[\hat\rho\hat D(\beta)
\left(-1\right)^{\hat a^{\dagger}\hat a}\hat D(-\beta)\right],
\end{equation}
where $\hat D(\beta)$ is the Weyl displacement operator with complex
argument $$\beta=\frac{1}{\sqrt2}(q+ip),$$ with $\hat a$ and $\hat
a^{\dagger}$ being the photon annihilation and creation operators.

Let us introduce the displaced density operator
\begin{equation}\label{eq.3}
\hat\rho_{\alpha}=\hat D^{-1}(\alpha)\hat\rho\hat D(\alpha).
\end{equation}
The Wigner function, which corresponds to the displaced density
operator, is of the form
\begin{equation}\label{eq.4}
W_{\hat\rho_{\alpha}}(q,p)= 2\,\mbox{Tr}\left[\hat\rho_{\alpha}\hat
D(\beta)\left(-1\right)^{\hat a^{\dagger}\hat a}\hat
D(-\beta)\right].
\end{equation}
By inserting the expression for the displaced density operator into
(\ref{eq.4}), one arrives at
\begin{equation}\label{eq.5}
W_{\hat\rho_{\alpha}}(q,p)= 2\,\mbox{Tr}\left[\hat
D^{-1}(\alpha)\hat\rho\hat D(\alpha)\hat D(\beta)
\left(-1\right)^{\hat a^{\dagger}\hat a}\hat D(-\beta)\right].
\end{equation}
In view of the properties of the Weyl displacement operator
\begin{eqnarray*}
&&\hat D(\beta)\hat D(\alpha)=\hat D(\beta+\alpha)\exp\Big[i\,
\mbox{Im}\,(\beta\alpha^*)\Big],\quad \hat D^{-1}(\alpha)=\hat
D(-\alpha),\\&&\hat D^{-1}(\alpha)\hat D^{-1}(\beta)= \Big(\hat
D(\beta)\hat D(\alpha)\Big)^{-1},\end{eqnarray*}
formula~(\ref{eq.5}) can be simplified as follows:
\begin{equation}\label{eq.6}
W_{\hat\rho_{\alpha}}(q,p)=
W_{\hat\rho}\Big(q+\sqrt{2}\,\mbox{Re}\,\alpha, ~p
+\sqrt{2}\,\mbox{Im}\,\alpha\Big).
\end{equation}
One can see that the Wigner function~(\ref{eq.4}) corresponding to
the displaced density operator is equal to the Wigner
function~(\ref{eq.2}) corresponding to the initial density operator
but with displaced arguments. The photon-number tomogram is the
photon distribution function (the probability to have $n$ photons)
in the state described by the displaced density operator
$\hat\rho_{\alpha}$~(\ref{eq.3}), i.e.,
\begin{equation}\label{eq.7}
\omega(n,\alpha)=P_n(\alpha)=\langle n |\hat\rho_{\alpha}| n
\rangle,\qquad n=0,1,2,\ldots
\end{equation}
As an example, we consider, following \cite{Mexica,LasResPNT},
the photon-number tomogram of the Gaussian state of one-mode
light described by the Wigner function of generic Gaussian form
\begin{equation}\label{eq.8}
W(q,p)=\frac{1}{\sqrt{\mbox{det}\,\sigma(t)}} \exp \left(-\frac
{1}{2} \mbox{\bf Q}\sigma^{-1}(t)\mbox{\bf Q}^T\right),
\end{equation}
where $\mbox{\bf Q}=\Big(p-\langle p\rangle,q-\langle q\rangle\Big)$
and the matrix $\sigma(t)$ is a real symmetric quadrature variance
matrix. The photon distribution function for one-mode mixed light
was obtained explicitly in terms of the Hermite polynomials of two
variables in \cite{PhysRev1}. The Hermite polynomials of two
variables $H^{\{\mbox {\bf R}\}}_{n_1\,n_2}(y_1,y_2)$, where
$n_1,n_2$ are nonnegative integers and $\mbox{\bf R}$ is a symmetric
2$\times$2 matrix, are determined by the generating function
\begin{eqnarray*}
&&\exp\left[-{1\over2}(x_1\,x_2)
\left(%
\begin{array}{cc}
  R_{11} & R_{12} \\
  R_{21} &R_{22} \\
\end{array}
\right)  \left(\begin{array}{cc} x_1\\
x_2\\\end{array}\right)+(y_1\,y_2)\left(\begin{array}{cc}R_{11}&R_{12}\\
R_{21}&R_{22}\\\end{array}\right) \left(\begin{array}{cc}x_1\\
x_2\end{array}\right)\right]=\\&&\sum_{n_1,n_2=0}^{\infty}\frac{x_1^{n_1}x_2^{n_2}}{n_1!n_2!}
\,H^{\{\mbox {\bf R}\}}_{n_1\, n_2}(y_1,y_2).
\end{eqnarray*}
Applying the scheme of calculations similar to the one used in
\cite{PhysRev1} to our photon-number tomogram~(\ref{eq.7}), we
arrive at the photon-number tomogram as a function of the Hermite
polynomial of two variables
\begin{equation}\label{eq.11}
\omega(n,\alpha)=\frac{P_0(\alpha)H^{\{\mbox {\bf R}\}}_{n\,n}
\Big(y_1(\alpha),y_2(\alpha)\Big)}{n!},
\end{equation}
where the matrix $\mbox {\bf R}$, which determines the Hermite
polynomial, reads
$$\mbox{\bf R}=\frac{1}{1+2T+4d}
\pmatrix{2\left(\sigma_{pp}-\sigma_{qq}-2i\sigma_{pq}\right) &1-4d
\cr 1-4d&2\left(\sigma_{pp}-\sigma_{qq}+2i\sigma_{pq}\right)}.$$
Here $d$ is the determinant of real symmetric quadrature variance
matrix $\sigma(t)$, i.e., $d=\sigma_{pp}\sigma_{qq}-\sigma_{pq}^2$
and $T$ is its trace $T=\sigma_{pp}+\sigma_{qq}.$ The arguments of
the Hermite polynomial are
\begin{eqnarray}
y_1(\alpha)=y_2^*({\alpha})&=&\frac{\sqrt{2}}{2T-4d-1}
\left[\left(\langle q\rangle-i\langle p\rangle+
\sqrt2\alpha^*\right)\left(T-1\right)\right.\nonumber\\
&&+\left.\left(\sigma_{pp}-\sigma_{qq}+2i\sigma_{pq}\right)\left(\langle
q\rangle+ i\langle p\rangle+\sqrt2\,\alpha\right)\right].
\end{eqnarray}
For the state with displaced Wigner function~(\ref{eq.4}), the
probability to have no photons $P_0(\alpha)$ reads
\begin{eqnarray}
P_0(\alpha)&=&\frac{2}{\sqrt L}\exp\left\{-\frac{1}{L}\left[
\left(2\sigma_{qq}+1\right)\left(\langle p\rangle
+\sqrt{2}\,\mbox{Im}\,\alpha\right)^2
+\left(2\sigma_{pp}+1\right)\left(\langle q\rangle+\sqrt{2}\,
\mbox{Re}\,\alpha\right)^2 \right]\right\}
\nonumber\\
&&\times\exp\left[\frac{4\sigma_{p q}}{L}\left(\langle p\rangle+
\sqrt{2}\,\mbox{Im}\,\alpha\right)\left(\langle q\rangle
+\sqrt{2}\,\mbox{Re}\,\alpha\right)\right],\label{eq.12a}
\end{eqnarray}
where $L=1+2T+4d$.

\section{CLASSICAL STATES}

Let us consider the classical state, for example, the state of a classical
particle with one degree of freedom with unit mass. We suppose that
the particle's position $q$ and momentum $p$ fluctuate due
to the interaction between the particle and some environment. The
particle's state is described by a probability distribution function
$f_{\rm cl}(q,p)$, which is nonnegative $f_{\rm cl}(q,p)\geq 0$ and
normalized. Here we assumed the normalization condition for
$f_{\rm cl}(q,p)$
\begin{equation}\label{eq.2}
\frac{1}{2\pi}\int f_{\rm cl}(q,p)\,dq\,dp = 1,
\end{equation}
analogously to the normalization condition of the Wigner function
$W(q,p)$.

The tomogram of this state is determined by the Radon
transform of the probability distribution function $f_{\rm cl}(q,p)$
\begin{equation}\label{eq.1a}
w_f\left (X,\,\mu ,\,\nu \right )=\int f_{\mbox{cl}}\left
(q,\,p\right )\, \delta \left (\mu q +\nu p -X\right )\,dq\,dp\,.
\end{equation}
It is called classical tomogram. The classical tomogram can be
determined as the expectation value of a delta-function calculated
with the help of the distribution function $f_{\rm cl}(q,p)$ in
the phase space
\begin{equation}\label{eq.1b} w_f\left (X,\,\mu ,\,\nu \right )=
\langle \delta\left (\mu q +\nu p -X\right )\rangle.
\end{equation}
The classical tomogram is nonnegative function and it is normalized
\begin{eqnarray*}
\int w_f\left (X,\,\mu ,\,\nu \right )\,d X=1.
\end{eqnarray*}
The physical meaning of the classical tomogram is that it is the
probability density for the particle's coordinate $X$, which is
measured in the phase-space reference frame subjected to the scaling
of axes and subsequent rotation with respect to the original
reference frame. In the same way as in the case of Fourier
transformation, where information contained in the function
is equivalent to information contained in its Fourier transform,
information on the particle's state contained in the distribution
function $f_{\rm cl}(q,p)$ is equivalent to information contained in its
Radon transform -- the classical tomogram $w_f\left (X,\,\mu ,\,\nu
\right )$. The Radon transformation is invertible
\begin{equation}\label{eq.1c}
f_{\mbox{cl}}\left (q,\,p\right)=\frac {1}{4\,\pi^2}\int w_f\left
(X,\,\mu ,\,\nu \right)\exp \left [-i\left (\mu q+\nu p -X\right )
\right ]\,dX\,d\mu \,d\nu,
\end{equation}
so the probability distribution function $f_{\rm cl}(q,p)$ can be
reconstructed in view of the classical tomogram
$w_f\left (X,\,\mu ,\,\nu \right)$.

The commutative star-product kernel
$K(X,\mu,\nu,X_1,\mu_1,\nu_1,X_2,\mu_2,\nu_2)$ for two classical
tomograms $w_{f1}(X_1,\mu_1,\nu_1)$ and $w_{f2}(X_2,\mu_2,\nu_2)$ is
\cite{MankoManko}
\begin{eqnarray*}
K_{\rm classic}(X,\mu,\nu,X_1,\mu_1,\nu_1,X_2,\mu_2,\nu_2) =
\frac{1}{(2\pi)^2}e^{i\left(X_1+X_2-X({\nu_1+\nu_2})/{\nu}\right)}
\delta\Big(\nu(\mu_1+\mu_2)-\mu(\nu_1+\nu_2)\Big).
\end{eqnarray*}
The relationship between tomographic star-product kernels in quantum
and classical mechanics reads \cite{Pilyavets}
\begin{eqnarray*}
K_{\mbox{\scriptsize{quant}}}(X,\mu,\nu,X_1,\mu_1,\nu_1,X_2,\mu_2,\nu_2)
=
K_{\mbox{\scriptsize{classic}}}(X,\mu,\nu,X_1,\mu_1,\nu_1,X_2,\mu_2,\nu_2)
e^{[i\left(\mu_2\nu_1-\mu_1\nu_2\right)/2]}.
\end{eqnarray*}

\section{CONCLUSIONS}

We reviewed two tomographic-probability approaches --- symplectic tomography
and photon-number tomography and showed that they are examples of the
star-product quantization scheme. We mentioned that, in the quantum
case, the state of the system can be described by the probability
distribution function instead of the density operator. This provides
the possibility to describe the objects of both the quantum and classical
natures using the same objects -- tomograms.

\section*{ACNOWLEDGMENTS}

The author thanks the Organizers of the Conference ''Foundation of
Probability and Physics 5'' and especially Prof. A. Khrennikov for
invitation and kind hospitality.

\end{document}